\documentclass[preprint,number,sort&compress]{elsarticle}

\usepackage{graphicx}
\usepackage{amsmath}
\usepackage{amssymb}
\usepackage{epstopdf}

\newcommand{\ud}{\mathrm{d}}

\begin{document}

\title{Noise-driven oscillations in microbial population dynamics}

\author[ediphys]{Bhavin S. Khatri\corref{cor1}\fnref{fn1}}
\ead{bhavin.khatri@physics.org}
\author[edibiol]{Andrew Free}
\ead{andrew.free@ed.ac.uk}
\author[ediphys]{Rosalind J. Allen}
\ead{rallen2@ph.ed.ac.uk}

\cortext[cor1]{Corresponding author}

\fntext[fn1]{Current address: Mathematical Biology Division, MRC National Institute for Medical Research, London, NW7 1AA, UK.}

\address[ediphys]{SUPA, School of Physics and Astronomy, The University of Edinburgh, Mayfield Road, Edinburgh EH9 3JZ, UK.}
\address[edibiol]{Institute of Evolutionary Biology, School of Biological Sciences, The University of Edinburgh, Mayfield Road, Edinburgh EH9 3JR, UK.}

\begin{abstract}
Microbial populations in the natural environment are likely to experience growth conditions very different from those of a typical laboratory experiment. In particular, removal rates of biomass and substrate are unlikely to be balanced under realistic environmental conditions. Here, we consider a single population growing on a substrate under conditions where the removal rates of substrate and biomass are not necessarily equal. For a large population, with deterministic growth dynamics, our model predicts that this system can show transient (damped) oscillations. For a small population, demographic noise causes these oscillations to be sustained indefinitely. These oscillations arise when the dynamics of changes in biomass are faster than the dynamics of the substrate, for example, due to a high microbial death rate and/or low substrate flow rates. We show that the same mechanism can produce sustained stochastic oscillations in a two-species, nutrient-cycling microbial ecosystem. Our results suggest that oscillatory population dynamics may be a common feature of small microbial populations in the natural environment, even in the absence of complex interspecies interactions.
\end{abstract}

\maketitle

\section{Introduction}

Microbial populations play an essential role in many important processes in the   natural and human environments, including nitrification of soil, carbon processing in ocean food chains and wastewater treatment. Extensive studies have informed our understanding of microbial population dynamics under standard conditions in the laboratory, yet outside the lab, different conditions may hold. In particular, nutrient supply may be unpredictable, nutrient and biomass removal rates may not balance, and population sizes may be small so that random fluctuations due to birth and death events play an important role. Recent studies have shown that microbial communities can undergo unpredictable divergence from similar initial conditions \cite{ourwork}, dramatic fluctuations in species composition  \cite{Fernandez1999} and even chaotic dynamics \cite{Becks2005,Becks2008,Beninca2008,Graham2007}.  Stochastic models for microbial ecosystem dynamics will be needed to obtain detailed understanding of such results. Here, we take a first step towards developing such models, by studying the effects of unbalanced biomass and substrate flow rates, and of demographic noise, on microbial population dynamics.

Microbial population dynamics are often studied in the context of the chemostat \cite{NOVICK1950,TheoryChemostat}: a well-stirred vessel in which  a population is maintained  under steady-state conditions, with constant inflow of substrate at fixed concentration, balanced by constant outflow of the vessel's contents (biomass and substrate). Assuming the standard Monod relation between growth rate and substrate concentration, the deterministic equations for biomass and substrate concentrations in the chemostat do not show either transient or sustained oscillations \cite{TheoryChemostat}. A large body of theoretical work has shown that oscillations can occur in extended versions of the classic chemostat equations, which include factors such as  time delays \cite{Bush1976,Xia2005}, feedback control \cite{Guo2009,Balakrishnan2002}, growth inhibition by substrate or products \cite{Hsu1992,Lenski1986}, age structure within the population \cite{Lemesle2008}, periodic nutrient input or washout rates \cite{Butler1985,Hale1983,Hsu1980}, and predator/prey interactions among populations within the chemostat \cite{Butler1983} - and indeed, oscillations have been observed in a number of chemostat experiments \cite{Hansen1980,Pirt1970,Porro1988}. However, almost all these models assume equal rates of substrate and biomass removal from the system (which is a consequence of the standard chemostat setup), and deterministic population dynamics. In this paper, we take a different approach: we use a very simple growth model, but take into account both imbalances in substrate and biomass flow rates, and fluctuations due to small population size.

For small microbial populations, individual birth and death events give rise to demographic fluctuations which may be significant compared to the total population size. For natural communities growing in microenvironments such as the interstices between soil grains or the surfaces of ocean particles, these demographic fluctuations may well be relevant. Moreover, recent work has shown that, in populations which expand to colonize new spatial territories, demographic fluctuations, due to small numbers at the boundary, can have dramatic consequences for the genetic structure of the population, even for large populations \cite{Hallatschek2007,Hallatschek2008}. From a theoretical point of view, it is well known that fluctuations can have important qualitative effects on the behaviour of dynamical systems  \cite{Pikovsky1997,Gammaitoni1998,McKane2005}.  In particular, Newman and McKane have shown recently that sustained oscillations can arise in stochastic dynamical systems driven by intrinsic noise, whose corresponding deterministic dynamical equations lead only to transient (damped) oscillations  \cite{McKane2005,Lugo2008,Boland2008}. This effect has been observed in a range of model systems including predator-prey models \cite{McKane2005,Lugo2008,Boland2008}, models for infection dynamics \cite{Alonso2007,Ghose2010}, cooperative games \cite{Galla2009,Bladon2010} and chemical reaction models for genetic and metabolic regulation \cite{McKane2007,Dauxois2009}. These oscillations arise because intrinsic noise excites the underlying damped oscillatory degrees of freedom of the system; they are self-sustaining, with all initial conditions leading to the same ensemble of stochastic orbits, and (similar to a limit-cycle) a characteristic amplitude and frequency are set by the parameters of the system.  In this paper, we show that the same mechanism can lead to sustained stochastic oscillations for a simple, one-population microbial growth model with unbalanced rates of substrate and biomass removal, and we suggest that it is likely to be a generic feature of microbial ecosystems with small population sizes \footnote{We note that this mechanism, in which stochastic oscillations are generated by intrinsic noise in the system dynamics, is distinct from the phenomenon of coherence resonance \cite{Pikovsky1997}, in which oscillations are caused by noise in an external driving force.}.

In Section \ref{sec:det}, we analyse a deterministic model for a single microbial population growing on a substrate, with unbalanced rates of biomass and substrate removal; we show that this system can show transient oscillations. In Section \ref{sec:stoch}, we show that demographic noise can cause sustained stochastic oscillations in this system. In Section \ref{sec:cycle}, we extend our analysis to a simple two-species, nutrient-cycling model.  Finally, we present our conclusions in Section \ref{sec:disc}.

\section{A single population: deterministic model}\label{sec:det}
We first consider, using deterministic equations, the dynamics of a single microbial population with biomass concentration $x(t)$ (units of microbial cells per litre), which consumes a substrate of concentration $s(t)$ (units of $\mu$M). Substrate flows into the system at a constant rate  $b$ (units of concentration/time) and is removed from the system with rate constant $R$ (units of time$^{-1}$), while biomass is removed with rate constant $D$ (also with units of time$^{-1}$). In a chemostat, the rates of substrate and biomass removal are equal ($R=D$), since the well-mixed contents of the chemostat are pumped out at a constant rate, and the microbial death rate is assumed to be negligible. In the natural environment, however, the situation is more complex: microbes may be subject to significant (e.g. phage-mediated) killing, substrate may be removed via consumption by competing organisms, and microbes may avoid being washed away by adhering to a surface. We do not therefore constrain the removal rates of substrate and biomass to be equal. Since in this section we use a deterministic approach, neglecting demographic fluctuations, our analysis is appropriate for a large population; in Section \ref{sec:stoch} we present contrasting results for a small population.

The equations governing the dynamics of our system are
\begin{eqnarray}
\frac{\ud x}{\ud t}\equiv \dot{x} &=& f(x,s) =  \mu(s) x- D x \label{Eq:ndot}\\
\frac{\ud s}{\ud t}\equiv \dot{s} &=& g(x, s) = -\gamma \mu(s) x + b - Rs \label{Eq:sdot}
\end{eqnarray}
where we assume that microbial growth is related to substrate concentration by the Monod function~\cite{Monod1949}\cite{Kovarova-Kovar1998}:
\begin{equation}
\left(\frac{\ud x}{\ud t}\right)_{\rm{growth}}=\mu(s)x \equiv \frac{vsx}{K+s}
\end{equation}
in which $v$ is the maximal growth rate and $K$ is the substrate concentration at which the growth rate is half-maximal. Thus the growth rate is linearly proportional to substrate concentration when $s\ll K$, but saturates at high substrate concentration $s\gg K$. The parameter $\gamma$ is the number of substrate molecules that need to be consumed to make one microbe (the inverse of the yield coefficient~\cite{GrowthBacCellBook}).

Eqs (\ref{Eq:ndot}) and (\ref{Eq:sdot}) have a single non-trivial fixed point (for which $\dot{x} = \dot{s} = 0$) at
\begin{eqnarray}\label{Eq:fpt}
 s^* &=& \frac{ K}{(v/D-1 )}\\\nonumber
x^* &=&  \frac{(b-Rs^*)}{\gamma D}
\end{eqnarray}

Starting from an arbitrary (nonzero) initial condition, the system will evolve towards this fixed point. To determine whether this happens monotonically or in an oscillatory manner, we analyse the system's dynamics close to the fixed point, by making the linear approximation \cite{Strogatz}:
\begin{equation}\label{Eq:jac}
\left(
             \begin{array}{c}
               \dot{\delta x} \\
               \dot{\delta s} \\
             \end{array}
           \right) = \left(
             \begin{array}{cc}
               \partial_{x} f & \partial_{s} f \\
               \partial_{x} g & \partial_{s} g \\
             \end{array}
           \right)_{(x^*,s^*)^T}\left(
             \begin{array}{c}
               \delta x \\
               \delta s \\
             \end{array}
           \right) \equiv \mathsf{J}^*\left(
             \begin{array}{c}
               \delta x \\
               \delta s \\
             \end{array}
           \right)
\end{equation}
where $\delta x=x-x^*$, $\delta s=s-s^*$, $\partial_z f$ and $\partial_s f$ are shorthand for $\partial f/\partial z$ and $\partial f/\partial s$ respectively, and $\mathsf{J}^*=\mathsf{J}(x^*,s^*)$ denotes the Jacobian matrix of first-order partial derivatives, evaluated at the fixed point $(x^*,s^*)$. If $\boldsymbol{u}_1$ and $\boldsymbol{u}_2$ are the eigenvectors of the Jacobian matrix $\mathsf{J}^*$ at the fixed point, then Eq.(\ref{Eq:jac}) implies that the system evolves as $\boldsymbol{u}_1 e^{\lambda_1 t} + \boldsymbol{u}_2 e^{\lambda_2 t}$, where $\lambda_1$ and $\lambda_2$ are the corresponding eigenvalues of $\mathsf{J}^*$. Real and negative eigenvalues of  $\mathsf{J}^*$ indicate exponential relaxation to the fixed point, while complex eigenvalues with a negative real part indicate exponentially decaying oscillations as the system approaches its fixed point \cite{Strogatz}. Using Eqs (\ref{Eq:ndot}), (\ref{Eq:sdot}) and (\ref{Eq:fpt}), we obtain:

\begin{eqnarray}\label{Eq:Jacobian}
\mathsf{J}^* &=&D\left(
             \begin{array}{cc}
               0 & \beta/\gamma \\
               -\gamma  & -(\beta+\chi) \\
             \end{array}
           \right)
\end{eqnarray}
where we have defined two dimensionless parameters:
\begin{equation}\label{Eq:Chi}
\chi \equiv \frac{R}{D}
\end{equation}
measures the rate of substrate removal relative to the rate of biomass removal, and
\begin{equation}\label{Eq:Beta}
\beta= \frac{\gamma x^*}{D}\left(\frac{\ud\mu(s)}{\ud s}\right)_{s=s^*}
\end{equation} measures the responsiveness of the microbial growth rate to changes in the substrate concentration. Increasing the substrate  inflow rate $b$ results in an increase in $\beta$ (since $x^*$ depends on $b$). The eigenvalues $\lambda$ of this Jacobian are given by:
\begin{equation}\label{Eq:EigenValue}
\frac{\lambda}{D}=\frac{-(\beta+\chi)\pm\sqrt{(\beta+\chi)^2-4\beta}}{2}
\end{equation}

\begin{figure}[h!]
\begin{center}
{\rotatebox{0}{{\includegraphics[width=0.7\textwidth]{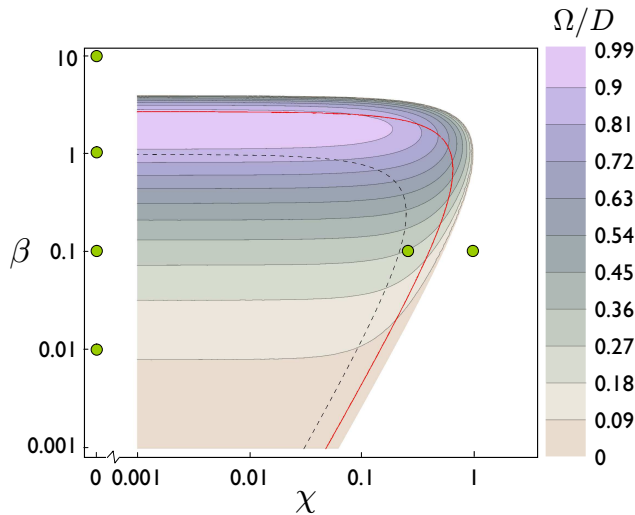}}}}
\caption{
Summary of the predicted behaviour of the single population model as a function of the parameter combinations $\beta$ and $\chi$. The coloured area shows the region of $\beta-\chi$ parameter space where our deterministic model predicts transient oscillations during relaxation to the steady state (i.e. the region where the eigenvalues of the Jacobian are complex with negative real part). Within this region, the contours show the frequency of the damped oscillations (Eqn.(\ref{Eq:OscFreq})). The solid red line shows the boundary of the region where our stochastic model predicts sustained oscillations (i.e. the values of $\beta$ and $\chi$ for which there is a peak in the biomass power spectrum Eqn.(\ref{Eq:nPSD})). The  dotted line represents the contour for which the quality factor $Q=1$ in Eqn.(\ref{Eq:q}). The green circles correspond to the parameter values used in the simulations of Figures \ref{Fig:DeterministicOsc}, \ref{Fig2} and \ref{Fig:StochTimeSeriesChi}, where the logarithmic $\chi$-axis is schematically extended to zero in order to represent simulations for $\chi=0$.} \label{Fig1}
\end{center}
\end{figure}

The case  $\chi \ge 1$ corresponds to a scenario where substrate is removed from the system faster than, or at the same rate as, biomass is removed. Under these circumstances,  the eigenvalues $\lambda$ are real and negative for any value of $\beta$, indicating that the system relaxes exponentially to its fixed point, with no oscillations (note that  $\chi = 1$ corresponds to the chemostat case \footnote{If $\chi =1$, Eqns. \ref{Eq:ndot} and \ref{Eq:sdot} share a symmetry which reduces the number of dynamical degrees of freedom to one, precluding the possibility of oscillating solutions.}). However, if substrate removal is slower than biomass removal ($\chi < 1$), then the eigenvalues $\lambda$ can be complex (with negative real part), implying that the system can  undergo transient oscillations in biomass and substrate concentration as it approaches the fixed point. These oscillations occur over a range of  parameter values corresponding to $(\beta+\chi)^2<4\beta$: i.e. for $\beta$ in the range  $\beta_-<\beta<\beta_+$, where $\beta_\pm=2-\chi\pm2\sqrt{1-\chi}$. The coloured region in Figure \ref{Fig1} shows the region of $\beta-\chi$ parameter space where transient oscillations are expected.

\begin{figure}[h!]
\begin{center}
{\rotatebox{0}{{\includegraphics[width=0.7\textwidth]{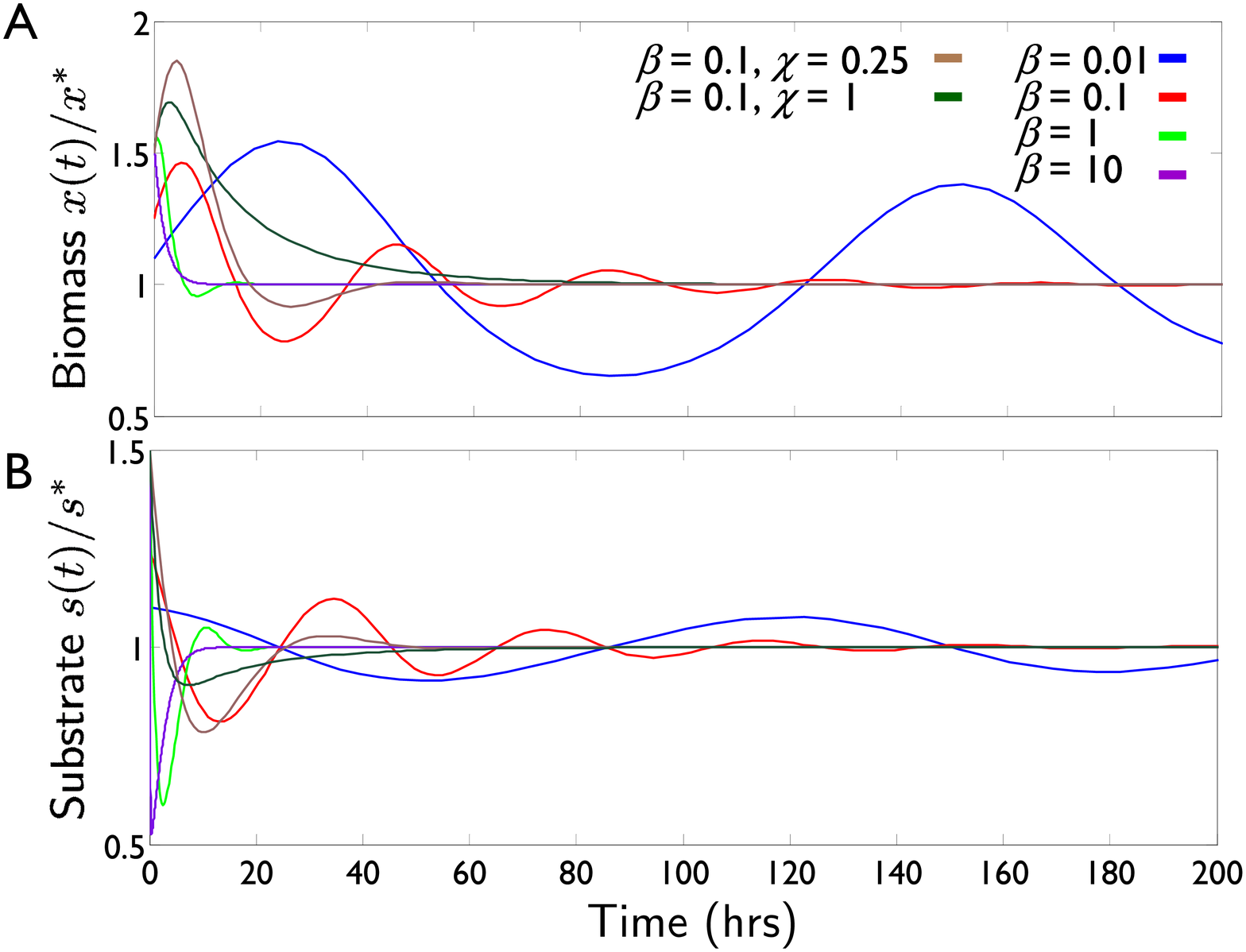}}}}
\caption{Deterministic time series for the growth of {\em{Escherichia coli}} on glucose, calculated by numerical simulation of the coupled ODEs represented by Eqns.(\ref{Eq:ndot})\&(\ref{Eq:sdot}), for parameters  $v=1\mathrm{hr}^{-1}$, $D=0.5\mathrm{hr}^{-1}$, $K=1\mu\mathrm{M}$, $\gamma=1.8\times 10^{10}$ substrate molecules. The 6 combinations of $\beta$ and $\chi$ correspond to the green circles in the $\beta-\chi$ phase diagram of Figure \ref{Fig1}, i.e. $\beta=\{0.01,0.1,1,10\}$ with $\chi=0$ ($b=\{0.01,0.1,1,10\}\mu\mathrm{Mhr}^{-1}$) and $\beta=0.1$ with $\chi=\{0.25,1\}$ ($b=0.225\mu\mathrm{Mhr}^{-1}$, $R=0.125\mathrm{hr}^{-1}$ and $b=0.6\mu\mathrm{Mhr}^{-1}$, $R=0.5\mathrm{hr}^{-1}$). Note that in the legend, $\chi=0$ unless otherwise stated.} \label{Fig:DeterministicOsc}
\end{center}
\end{figure}

To show that these transient oscillations are relevant for microbial populations, we plot in  Figure \ref{Fig:DeterministicOsc} simulated dynamical trajectories for biomass and substrate concentration, for a parameter set whose values are chosen  to  correspond approximately to \textit{Escherichia coli} growing on glucose: $v=1\mathrm{hr}^{-1}$, $D=0.5\mathrm{hr}^{-1}$, $K=1\mu\mathrm{M}$ and $\gamma=1.8\times 10^{10}$ substrate molecules consumed to produce one microbe \cite{Kovarova-Kovar1998}\cite{GrowthBacCellBook}\cite{Cohen1954}. Keeping these parameters fixed, the dimensionless parameters $\beta$ and $\chi$ in Eqs. (\ref{Eq:Chi}) and (\ref{Eq:Beta}) are controlled by the substrate flow rates with the following numerical values: $\beta=b-R$ and $\chi=2R$; we choose $(\beta,\chi)$ combinations  corresponding to the green circles in the phase diagram of Figure \ref{Fig1}. Significant transient oscillations are indeed observed in our simulations (on a timescale of typically tens of hours) for those parameter combinations which lie inside the coloured region of Figure \ref{Fig1}.

Our analysis of the eigenvalues of the Jacobian (Eqn.(\ref{Eq:EigenValue})) also provides information on the frequency $\Omega$ of the transient oscillations that occur during relaxation to the fixed point. This is given by the imaginary part of the eigenvalues $\lambda$:
\begin{equation}\label{Eq:OscFreq}
\Omega=\frac{D}{2}\sqrt{4\beta-(\beta+\chi)^2}.
\end{equation}
The predicted frequency $\Omega$ is indicated by the contours in the phase diagram of Figure \ref{Fig1}. Increasing $\chi$ -- i.e. either increasing the rate of substrate  removal or decreasing the rate of biomass removal -- decreases the oscillation frequency, while the dependence on $\beta$ is nonmonotonic, with a peak in $\Omega$ for intermediate values of $\beta$ (for fixed $\chi$).

\begin{figure}[h!]
\begin{center}
{\rotatebox{0}{{\includegraphics[width=0.7\textwidth]{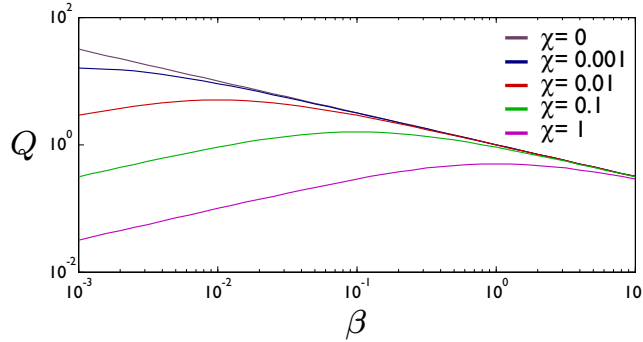}}}}
\caption{Quality factor $Q$, predicted by Eq.(\ref{Eq:q}), plotted as a function of $\beta$ for  various values of $\chi$.
 \label{newfig2}}
\end{center}
\end{figure}

To understand better the nature of the oscillations, we also compute the quality factor $Q$, which is a dimensionless measure of the extent to which oscillations are damped, and corresponds roughly to the number of oscillations that occur before the oscillations die away in the deterministic time series (Fig. \ref{Fig:DeterministicOsc}). For a simple harmonic oscillator, $Q$ is the ratio of the energy stored to energy dissipated over an oscillation cycle. As we show in \ref{app:qfactor}, close to the fixed point, our deterministic model can be mapped onto the equations for a simple harmonic oscillator; this allows us to approximate the $Q$-factor of the oscillations as
\begin{equation}\label{Eq:q}
Q\approx\frac{\sqrt{\beta}}{\beta+\chi}.
\end{equation}

Figure \ref{newfig2} shows $Q$ (from Eq.(\ref{Eq:q})) as a function of $\beta$, for several values of $\chi$. If $\beta > \chi$, the quality factor of the oscillations increases as $\beta$ decreases: reduced inflow of substrate (keeping other parameters fixed) will lead to more pronounced oscillations. However, if $\beta<\chi$, the opposite scenario holds; the quality factor of the oscillations will increase on increasing $\beta$ - i.e. on increasing the inflow of substrate, keeping all other parameters fixed. The dashed black line in the phase diagram of Figure \ref{Fig1} shows the contour of $\beta$ and $\chi$ for which $Q=1$; inside this line we expect to see that oscillations persist for a long time, while outside this line, oscillations, even though present, are strongly  damped and rapidly decay. To understand this further, we  rescale the substrate deviation from the fixed point as $\delta s'=\frac{\beta+\chi}{\gamma}\delta s$ and scale time by $D$, the death/removal rate of biomass. In these scaled units, the substrate relaxes like $\dot{\delta s'}=-(\beta+\chi)(\delta x +\delta s')$ and the biomass varies as $\dot{\delta x}= \frac{\beta}{\beta+\chi}\delta s'$. We can then identify the relaxation rate constant of the substrate as $\lambda_s=\beta+\chi$ and that for the biomass as $\lambda_x=\frac{\beta}{\beta+\chi}$, from which we see that
\begin{equation}\label{Eq:lambdaQ}
\lambda_x/\lambda_s=Q^2.
\end{equation}
In other words the more quickly the biomass responds to changes in substrate relative to the rate of relaxation of the substrate the higher the quality factor of the oscillations and the longer they will persist. Conversely, if the substrate relaxes more quickly than the biomass can respond, we see that the quality factor $Q\ll1$ and the oscillations will be very strongly damped (for those values of $\beta$ and $\chi$ expected to give rise to oscillations - as shown in Fig.\ref{Fig1}).

\begin{figure}[h!]
\begin{center}
{\rotatebox{0}{{\includegraphics[width=0.7\textwidth]{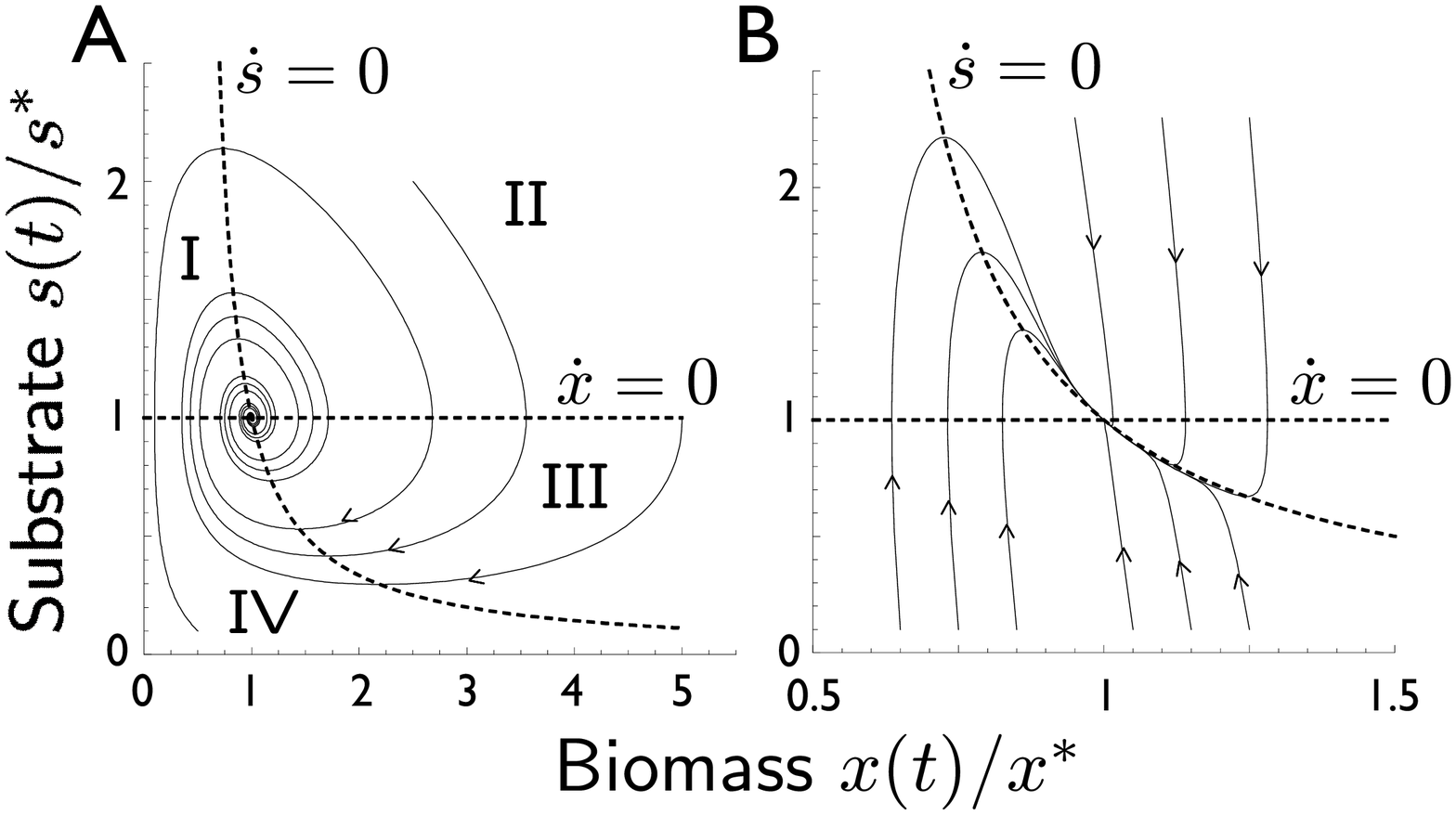}}}}
\caption{Phase plane plots of the dynamics of microbial growth for a small substrate influx rate ($\beta=0.1$) that gives rise to  damped oscillations (A) and for a large substrate influx rate ($\beta=10$) that gives exponential relaxation (B) to steady state (in both cases $\chi=0$). In panel A ($\beta=0.1$) oscillations arise due to a cyclical imbalance of growth vs death and net inflow vs consumption. In contrast, in panel B ($\beta=10$) the substrate relaxes quickly to give a quasi-equilibrium of inflow vs consumption; biomass and substrate subsequently evolve slowly  to the fixed point.} \label{Fig:PhasePlane}
\end{center}
\end{figure}

The underlying cause of the transient oscillations can be understood in terms of  changes in balance between microbial growth and death and between consumption and net inflow of substrate. The phase plane plot of Fig.~\ref{Fig:PhasePlane}A shows how the biomass and substrate concentrations change during the approach to the fixed point, for trajectories starting from three different initial conditions, with  $\beta=0.1$ and $\chi=0$. Transient oscillations are apparent from the fact that the  trajectories spiral into the fixed point.  The nullclines $\dot{x}=0$ and $\dot{s}=0$ ($s=s^*$ and $b-Rs=\gamma\mu(s)x$ respectively, obtained by setting Eqns.(\ref{Eq:ndot}) and (\ref{Eq:sdot}) to zero), shown as dashed lines in Fig.~\ref{Fig:PhasePlane}A, partition the phase plane into regions where $\dot{x}$ and $\dot{s}$ have different sign combinations. In region I, the rate of biomass growth exceeds the rate of death ($\dot{x}>0$), and the rate of substrate inflow exceeds its rate of consumption and removal ($\dot{s}>0$); thus both biomass and substrate concentrations increase. However, eventually the biomass concentration becomes high enough that the rate of substrate consumption (combined with its outflow) exceeds its rate of inflow; the system then enters region II where the substrate concentration is decreasing ($\dot{s}<0$), but remains high enough that biomass continues to accumulate  ($\dot{x}>0$). When the substrate concentration decreases to the point that the biomass growth rate is less than its net loss rate, the system enters region III: here the biomass concentration decreases  ($\dot{x}<0$), and the substrate concentration continues to decrease  ($\dot{s}<0$). When the biomass concentration has decreased sufficiently, however, the rate of substrate consumption becomes low enough that the substrate inflow rate exceeds its net loss rate, and the substrate concentration starts to increase ($\dot{s}>0$), while the biomass concentration continues to decrease  ($\dot{x}<0$); this corresponds to region IV in Fig.~\ref{Fig:PhasePlane}A. Finally, the increase in substrate concentration causes the the biomass growth rate to increase, such that eventually the biomass growth rate exceeds its net loss rate, and the system again enters region I. This oscillation mechanism relies on the biomass relaxing more quickly or on a similar timescale to the substrate concentration ($\lambda_x\ge \lambda_s$); rapid changes in biomass concentration lead to deficits or excesses in the substrate concentration, which are only slowly restored towards equilibrium by the flow of substrate into or out of the system.

In contrast, Fig.~\ref{Fig:PhasePlane}B shows phase plane trajectories for a case where our analysis does not predict oscillations ($\beta=10$, $\chi=0$). In this case, when $\beta+\chi\gg 1$ the substrate concentration relaxes much more quickly than the biomass concentration ($\lambda_s\gg\lambda_x$). Starting from given initial substrate and biomass concentrations, first the substrate concentration adjusts such that its inflow and net loss rates are equal (i.e. $\dot{s}\approx0$), for the given biomass concentration, then the biomass concentration slowly relaxes to the fixed point (making the quasi-equilibrium assumption that $\dot{\delta s'}\approx 0$, we find $\dot{\delta x}\approx -\lambda_x\delta x$), with accompanying changes in the substrate concentration. In other words, the trajectories rapidly approach the nullcline $\dot{s}=0$, then more slowly move along this nullcline to the fixed point.

\section{A single population: stochastic model}\label{sec:stoch}

We next consider the dynamics for the same model, but for small microbial populations. In this case, randomness in the birth and death/removal of individual microbes gives rise to stochastic fluctuations (or ``demographic noise''), which cannot be neglected. McKane and Newman \cite{McKane2005} have shown that, for a simple two-species predator-prey ecosystem,  this demographic noise can  produce sustained stochastic oscillations where the equivalent deterministic dynamical system shows only transient oscillations. Here, we show that the same effect happens in our single-population model.

Our system can be described by the following chemical reaction scheme:
\begin{eqnarray}
&&\emptyset \xrightarrow{} S\label{eq:reacs1}\\
&&X+\gamma S \xrightarrow{} 2X\label{eq:reacs2}\\ &&X \xrightarrow{} \emptyset\label{eq:reacs3}\\ &&S \xrightarrow{} \emptyset \label{eq:reacs4}
\end{eqnarray}
in which $X$ and $S$ denote  microbes and substrate molecules respectively;  we denote the numbers of microbes and substrate molecules as $n_X$ and  $n_S$ respectively.  Substrate molecules enter the system at a constant rate $b'$ (Eq.\ref{eq:reacs1}). Microbes replicate upon consuming $\gamma$ molecules of substrate (Eq.\ref{eq:reacs2}), at a rate given by the Monod function $\mu'(n_S)n_X =\frac{v n_S}{K'+n_S} n_X$. In addition, microbes are removed from the system at rate $Dn_X$ (Eq.\ref{eq:reacs3}) and substrate molecules are removed at rate $Rn_S$ (Eq.\ref{eq:reacs4}). The new parameters $b'$ and $K'$ arise because our units are now absolute numbers of microbes and substrate molecules rather than concentrations as in Eqs.(\ref{Eq:ndot})  and (\ref{Eq:sdot}); these parameters can easily be related to $b$ and $K$ in our deterministic model \footnote{For example  $b'$ (in moles per hour) is given by $bV$ where $b$ is measured in moles per litre per hour and the volume of the system $V$ is in litres.}. The parameters $v$, $R$, $D$ and $\gamma$, which do not depend on the units of substrate and biomass, are the same as in our deterministic model.  We assume that reactions (\ref{eq:reacs1}) to (\ref{eq:reacs4}) are Poisson processes - i.e. they happen randomly in time with average rates determined by their rate constants.

We use stochastic simulations to generate dynamical trajectories corresponding to the reaction scheme (\ref{eq:reacs1}) - (\ref{eq:reacs4}). In principle, this could be done using a kinetic Monte Carlo scheme such as the Gillespie algorithm \cite{Gillespie1977}, in which a single reaction happens in each timestep. However this would be extremely inefficient because the number of substrate molecules is typically very much larger than the number of microbes, so that reactions (\ref{eq:reacs1}) and (\ref{eq:reacs4}) would happen much more often than reactions  (\ref{eq:reacs2}) and (\ref{eq:reacs3}). Instead, we map the system onto a set of differential equations for the concentrations of biomass and substrate, as in Section \ref{sec:det}, but including a stochastic noise term which accounts for the effects of fluctuations (i.e. a set of Langevin equations for the biomass and substrate concentrations). This is done using  a Kramers-Moyal expansion of the Master Equation corresponding to Eqs (\ref{eq:reacs1}) - (\ref{eq:reacs4}), to obtain a non-linear Fokker-Planck equation, which is an approximate description of the stochastic dynamics. We then write down the equivalent non-linear Langevin equation whose stochastic trajectories correspond to this Fokker-Planck equation \cite{vanKampen,Gillespie1977}. Carrying out this procedure, as detailed in \ref{app:kramers-moyal}, we find
\begin{equation}\label{Eq:Kramers-Moyal_Langevin_conc}
\frac{\ud\boldsymbol{\phi}}{\ud t}= \mathsf{A}(\boldsymbol{\phi}) +\mathsf{B}^{1/2}(\boldsymbol{\phi})\boldsymbol{\xi}(t)
\end{equation}
where $\boldsymbol\phi=(x,s)^T$ is the vector of concentrations and the noise vector $\boldsymbol{\xi}(t)$ is Gaussian with zero mean and moment $\langle\boldsymbol{\xi}\boldsymbol{\xi}^T\rangle=\mathsf{I}\delta(t-t')/V$, where $\mathsf{I}$ is the identity matrix. The vector $\mathsf{A}$, given by
\begin{equation}\label{Eq:AVector_conc}
\mathsf{A}=\left(
             \begin{array}{c}
               x\mu(s)-Dx \\
               -\gamma x \mu(s) +b -Rs \\
             \end{array}
           \right)
\end{equation}
describes the deterministic time evolution of the system, as in Eqs.(\ref{Eq:ndot}) and (\ref{Eq:sdot}) ($\mu(s) = \frac{vs}{K+s}$ being the Monod growth function). The matrix $\mathsf{B}$, which ensures the correct coupling between biomass and substrate fluctuations, is given by
\begin{equation}\label{Eq:Bmatrix_conc}
\mathsf{B}=\left(
             \begin{array}{cc}
               x\mu(s)+Dx & -\gamma x \mu(s) +Dx\\
               -\gamma x \mu(s) & \gamma^2 x \mu(s) + b + Rs \\
             \end{array}
           \right)
\end{equation}
We generate trajectories corresponding to this Langevin equation using an Euler integration scheme.

\begin{figure}[h!]
\begin{center}
{\rotatebox{0}{{\includegraphics[width=.7\textwidth]{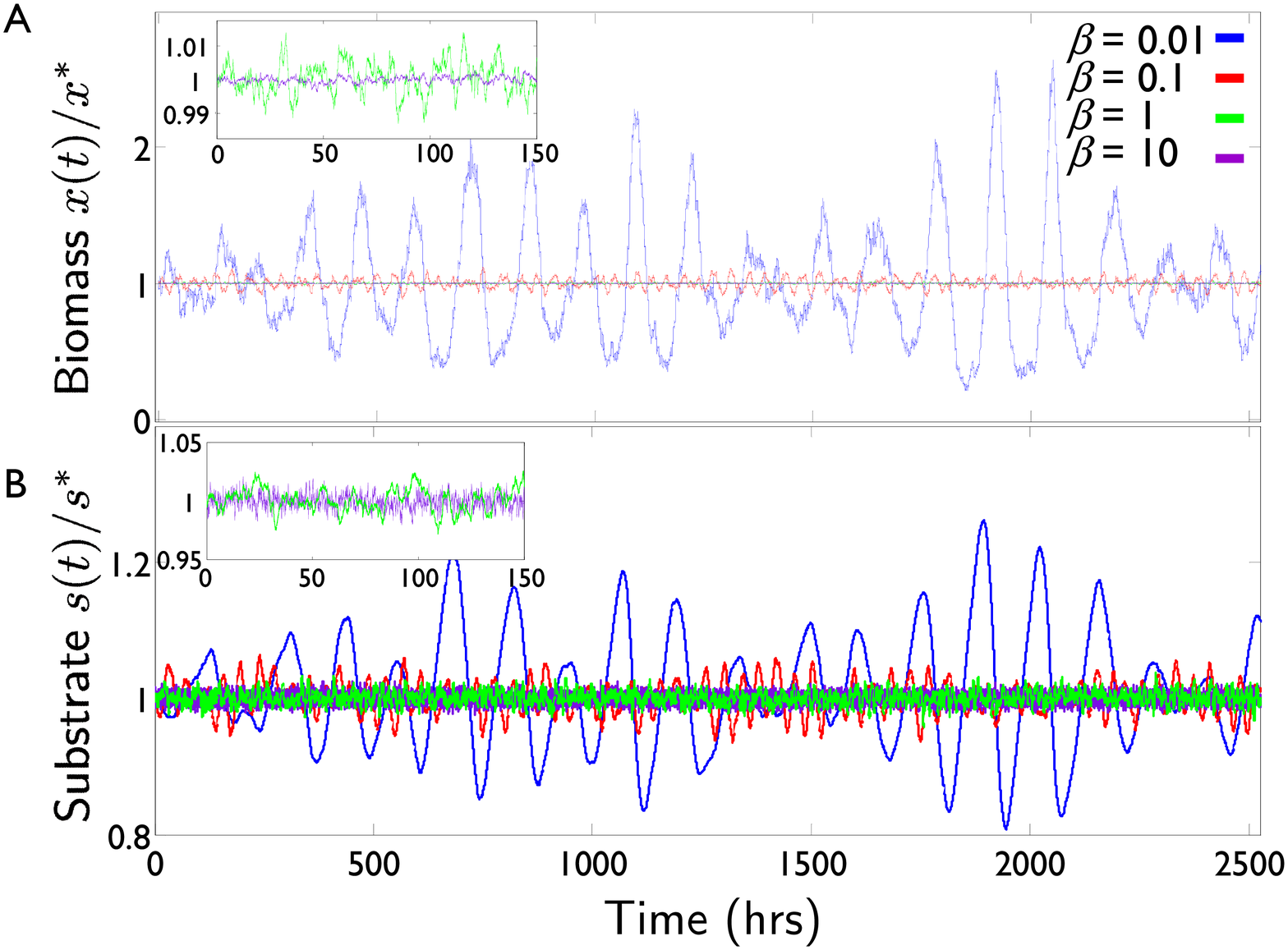}}}}
\caption{Stochastic time series for the growth of {\em{Escherichia coli}} on glucose, obtained by numerical simulation of the Langevin Equations (\ref{Eq:Kramers-Moyal_Langevin_conc}), for parameters  $v=1\mathrm{hr}^{-1}$, $D=0.5\mathrm{hr}^{-1}$, $K=1\mu\mathrm{M}$, $\gamma=1.8\times 10^{10}$ substrate molecules. Panel A shows the biomass concentration, normalized by its average, while panel B shows the normalized substrate concentration. Results are shown for $R=0$ and $b=0.01$, $0.1$, $1$ and $10$ $\mu$Mhr$^{-1}$, corresponding to $\chi=0$, $\beta= \{0.01,0.1,1,10\}$ (these parameter combinations are shown as green circles in the $\beta-\chi$ phase diagram of Figure \ref{Fig1}). The insets show the fluctuations of biomass (A) and substrate (B) for $\beta=1$ and $\beta=10$, in more detail and on a shorter timescale. The colour codes in panels A and B are the same.
 \label{Fig2}}
\end{center}
\end{figure}

\begin{figure}[h!]
\begin{center}
{\rotatebox{0}{{\includegraphics[width=.7\textwidth]{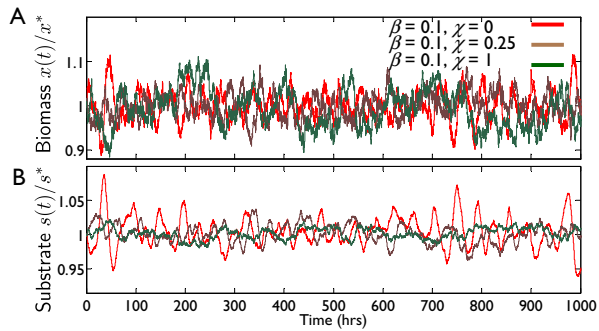}}}}
\caption{Stochastic time series for the growth of {\em{Escherichia coli}} on glucose, obtained by numerical simulation of the Langevin Equations (\ref{Eq:Kramers-Moyal_Langevin_conc}), for parameters  $v=1\mathrm{hr}^{-1}$, $D=0.5\mathrm{hr}^{-1}$, $K=1\mu\mathrm{M}$, $\gamma=1.8\times 10^{10}$ substrate molecules. Panel A shows the biomass concentration, normalized by its average, while panel B shows the normalized substrate concentration. Results are shown for  $\beta=0.1$ and $\chi = \{0,0.25,1\}$ (green circles in Figure \ref{Fig1}). These parameter combinations correspond to $b=0.1\mu$M/hr, $R=0$ (red line), $b=0.225\mu$M/hr, $R=0.125$hr$^{-1}$ (brown line) and $b=0.6\mu$M/hr, $R=0.5$hr$^{-1}$ (green line) respectively. The colour codes in panels A and B are the same.
 \label{Fig:StochTimeSeriesChi}}
\end{center}
\end{figure}

Figures \ref{Fig2} \& \ref{Fig:StochTimeSeriesChi} show the resulting stochastic dynamical trajectories, for the same parameter set used for the deterministic trajectories of Figure \ref{Fig:DeterministicOsc}, representing {\em{Escherichia coli}} growing on glucose, for a  system volume $V=1\mathrm{ml}$. Trajectories are plotted for combinations of $b$, $R$ and $D$ corresponding to the green circles in the phase diagram of  Figure \ref{Fig1}.  The average number of substrate molecules represented by each of these simulations is $\approx1$nmol and the average number of microbes varies from $\approx60,000$ at $b=10\mu$M/hr to $\approx60$ at $b=0.01\mu$M/hr (taking $R=0$): the microbial population is indeed small enough (especially at low substrate inflow rates) that we would expect  demographic fluctuations to play a significant role. Figures \ref{Fig2} \& \ref{Fig:StochTimeSeriesChi} show the microbial biomass and substrate concentrations (panels A and B respectively),  normalised by their steady-state time average.  For the two parameter combinations which lie outside the region of predicted oscillations in Figure \ref{Fig1} -- a high rate of substrate influx ($\chi=0,\beta=10$, Figure \ref{Fig2}) or rapid substrate removal rate ($\chi=1, \beta=0.1$, Figure \ref{Fig:StochTimeSeriesChi}) --  the stochastic simulations show random fluctuations about the steady state with no tendency to oscillate. However, for all other parameter combinations, which lie inside the region of predicted oscillations in Figure \ref{Fig1}, the stochastic simulations show sustained oscillations. Comparing the deterministic trajectories of Figure \ref{Fig:DeterministicOsc} with the stochastic trajectories of Figure \ref{Fig2} \& \ref{Fig:StochTimeSeriesChi}, we see that, at least for these parameter combinations, sustained oscillations in the stochastic system occur for parameter sets where the deterministic system shows transient oscillations. Following McKane and Newman \cite{McKane2005}, we reason that these sustained stochastic oscillations are generated by the continuous excitation of the oscillatory modes of the system by the intrinsic demographic fluctuations.

\begin{figure}[h!]
\begin{center}
{\rotatebox{0}{{\includegraphics[width=0.7\textwidth]{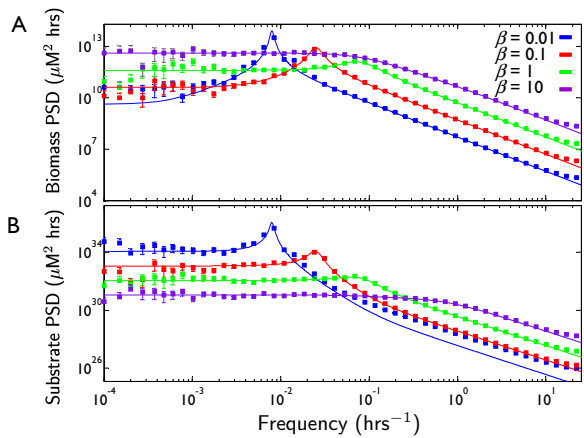}}}}
\caption{Simulation (points) and analytical (solid lines) results for the power spectrum of biomass (A) and substrate (B) concentration fluctuations, for substrate influxes $b$ of 0.01, 0.1, 1 and 10 $\mu$Mhr$^{-1}$, corresponding to $\beta=\{0.01,0.1,1,10\}$, with $\chi=0$ in all cases. Clear peaks in the power spectrum for $\beta=\{0.01,0.1,1\}$ indicate the presence of sustained stochastic oscillations. Note that the colour codes in panels A and B are the same.
 \label{Fig3}}
\end{center}
\end{figure}

\begin{figure}[h!]
\begin{center}
{\rotatebox{0}{{\includegraphics[width=0.7\textwidth]{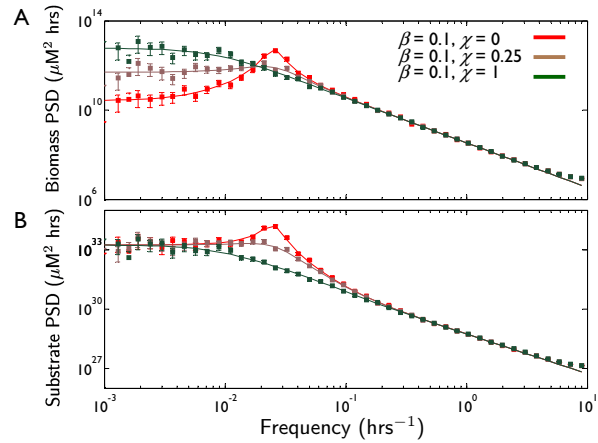}}}}
\caption{Simulation (points) and analytical (solid lines) results for the power spectrum of biomass (A) and substrate (B) concentration fluctuations, for $\beta=0.1$ and $\chi=\{0,0.25,1\}$, corresponding to the $(b,R)$ combinations ($0.1\mu$M/hr, 0), ($0.225\mu$M/hr, $0.125$hr$^{-1}$) and ($0.6\mu$M/hr, $0.5$hr$^{-1}$), respectively.  Clear peaks in the power spectrum for $\chi=\{0,0.25\}$ indicate the presence of sustained stochastic oscillations. Note that the colour codes in panels A and B are the same.
 \label{Fig:PSD_Chi}}
\end{center}
\end{figure}

To analyse in more detail the nature of the oscillations, we plot in Figures \ref{Fig3} and \ref{Fig:PSD_Chi} (squares) the power spectrum of biomass and substrate concentration fluctuations, for stochastic simulations with the same parameters sets as in Figures \ref{Fig2} and \ref{Fig:StochTimeSeriesChi}. Assuming a stationary stochastic process\footnote{A stationary stochastic process is defined to be one whose properties do not change with time and for which the autocorrelation function only depends on the difference between two time points $\tau$ and not on their absolute times.}, the Wiener-Khinchin theorem \cite{vanKampen}, relates the power spectral density $I(\omega)$ to the Fourier Transform of the autocorrelation function of the concentration fluctuations in the steady state -- for example for the biomass concentration
\begin{equation}
I_x(\omega) = \int_{-\infty}^{\infty} \langle\delta x(t) \delta x(t+\tau)\rangle e^{i \omega \tau} d\tau
\end{equation}
where we define the concentration fluctuation $\delta x(t)$ relative to the average $\langle x \rangle$: $\delta x(t) \equiv (x(t) - \langle x \rangle)$. An equivalent formula holds for the power spectral density $I_s(\omega)$ of the substrate concentration fluctuations. A peak in the power spectral density at a given frequency is a signature of sustained oscillations at that frequency in a noisy time series. Figures \ref{Fig3} and \ref{Fig:PSD_Chi} clearly show that  sustained oscillations indeed arise for those parameter sets for which the deterministic model produces transient oscillations (see Figures \ref{Fig1}, \ref{Fig:DeterministicOsc} and \ref{Fig:PhasePlane}).

We can also obtain analytical predictions for the power spectral density, using the van Kampen system size expansion \cite{vanKampen} of the Master Equation corresponding to Eqs (\ref{eq:reacs1}) to (\ref{eq:reacs4}). This procedure, the details of which are given in \ref{app:vk}, results in the following expressions for the power spectral densities of the biomass and substrate oscillations:

\begin{equation}\label{Eq:nPSD}
I_x(\omega)=\frac{\Lambda}{\gamma}\frac{(\beta+\chi)^2+\chi^2+2\omega^2/D^2 }{(\beta -\omega^2/D^2)^2+(\beta+\chi)^2\omega^2/D^2}
\end{equation}

\noindent and

\begin{equation}\label{Eq:sPSD}
I_s(\omega)=\Lambda\gamma \frac{2+ \omega^2/D^2}{(\beta-\omega^2/D^2)^2+(\beta+\chi)^2\omega^2/D^2}.
\end{equation}

\noindent where $\Lambda=\beta(V\partial_s\mu|_{s=s^*})^{-1}$. Eqs (\ref{Eq:nPSD}) and (\ref{Eq:sPSD}), which are  plotted in  Figures \ref{Fig3} and \ref{Fig:PSD_Chi} (solid lines) are in excellent agreement with the simulation results \footnote{In Figure \ref{Fig3}, the deviation at high frequency between the analytical and simulation results for the substrate power spectrum for small $\beta$ (corresponding to low substrate inflow rate)  may be attributed to the fact that here  the relative fluctuations in biomass and substrate are both roughly of order $\sim 1$, so that one would expect the Kramers-Moyal expansion used to obtain the simulation algorithm to give different results from the van Kampen expansion used to obtain the expression for the power spectrum.}. Note that the amplitude of the PSD varies as $V^{-1}$, which indicates  that the size of the fluctuations decrease as the volume of the system increases, as expected. These analytical results also allow us to determine which combinations of $\beta$ and $\chi$ give rise to a peak in the power spectrum, and hence to sustained stochastic oscillations. The red line in  Fig.~\ref{Fig1} shows the region of the $\beta$-$\chi$ parameter space in which sustained stochastic oscillations are expected from Eqs (\ref{Eq:nPSD}) and (\ref{Eq:sPSD}). Interestingly, this region lies {\em{inside}} the region where transient oscillations are predicted for the deterministic model. This suggests that, at least for this system,  the presence of transient oscillations in the deterministic model is a necessary, but not sufficient, condition, for the presence of sustained oscillations in the stochastic model.

As well as predicting the presence or absence of stochastic oscillations for a given parameter set, it is also important to know how pronounced these oscillations are expected to be. For example, in Figure \ref{Fig2}, oscillations are present both for $\beta=0.1$ (red line) and $\beta=0.01$ (blue line), but they are much more pronounced in the latter case. Similarly, for $\beta=0.1$ and $\beta=0.01$ we see a corresponding increase in sharpness of the peaks in the power spectra. We can understand this in terms of changes in the quality factor of the oscillations examined in Section \ref{sec:det}. As $\beta$ is decreased (for $\chi=0$) the ratio of the timescales of biomass and substrate relaxation decreases, leading to an increase in the quality factor as predicted by Eq.(\ref{Eq:lambdaQ}). More generally, as Fig.\ref{newfig2} and Eq.(\ref{Eq:q}) shows, the condition $\beta+\chi\ll\sqrt{\beta}$ must be satisfied to show significant oscillations; for example, in Fig.\ref{Fig:PSD_Chi}, we see that for $\beta=0.1$ the peak in the power gradually disappears as $\chi$ is increased due to a decrease in the relaxation time of the substrate relative to biomass.

\section{A two-species nutrient-cycling ecosystem}\label{sec:cycle}

Our results thus far demonstrate that demographic noise can result in sustained oscillations for a single microbial population whose rates of biomass and substrate removal are not balanced. In the natural environment, however, the waste product of one microbial population may form the substrate for another, leading to  microbial ecosystems with complex webs of crossfeeding interactions. Would we expect to see oscillations in such complex ecosystems?

\begin{figure}[h!]
\begin{center}
{\rotatebox{0}{{\includegraphics[width=0.7\textwidth]{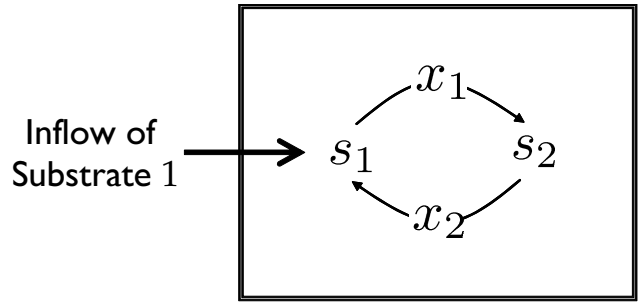}}}}
\caption{Schematic illustration of a two-species nutrient-cycling ecosystem.
 \label{Fig:cycle}}
\end{center}
\end{figure}

Figure \ref{Fig:cycle} shows a schematic illustration of a simple nutrient-cycling microbial ecosystem. Here, population 1 consumes substrate 1 and produces substrate 2; this is in turn consumed by a second  population, which releases as its waste product substrate 1. We assume that only substrate 1 is supplied by the external environment. This model might represent for example the cycling of carbon between methane and carbon dioxide by methanogens and methanotrophs, or the cycling of sulphur between sulphide and sulphate by sulphur oxidising and sulphur reducing bacteria; here, the environment is assumed to supply one of the forms of carbon/sulphur at a constant rate, and other necessary inputs (e.g. oxygen  for the oxidation reaction and hydrogen for the reduction reaction) are assumed to be available in plentiful supply.   The cycle of Figure \ref{Fig:cycle} can be represented by the following set of deterministic equations:

\begin{eqnarray}\label{Eq:cycle}
\frac{\ud x_1}{\ud t} &=& \frac{v x_1 s_1 }{K+s_1} - Dx_1\\
\nonumber\frac{\ud x_2}{\ud t} &=& \frac{v x_2 s_2}{K+s_2} - Dx_2\\
\nonumber\frac{\ud s_1}{\ud t} &=& -\gamma \frac{v x_1s_1}{K+s_1} + f \gamma \frac{v x_2s_2}{K+s_2}+b\\
\nonumber\frac{\ud s_2}{\ud t} &=& -\gamma \frac{v x_2s_2}{K+s_2} + f \gamma \frac{v x_1s_1}{K +s_1}
\end{eqnarray}
Here, $x_1$ and $x_2$ are the biomass concentrations of populations 1 and 2, while $s_1$ and $s_2$ are the concentrations of the two substrates. The factor $f$ arises in the equations for $\ud s_1/\ud t$ and $\ud s_2/\ud t$ because we suppose that for every gamma substrate molecules which they consume, microbes use a fraction $(1-f)$ for growth and excrete the remaining $f$ in the form of waste product, which ensures conservation of matter. We assume for simplicity that both populations have the same growth/removal parameters $v$, $K$, $\gamma$ and $D$, and that the substrate removal rate $R$ is zero. The inflow rate of substrate 1 is given by $b$. Eqs.(\ref{Eq:cycle}) have a single non-trivial fixed point with steady state biomass and substrate concentrations given by
\begin{eqnarray}\label{Eq:MutualFixedPoint}
x_1^* &=&  \frac{b}{\gamma D(1-f^2)}\\
\nonumber x_2^* &=&  \frac{b f}{\gamma D(1-f^2)}\\
\nonumber s_1^*=s_2^* &=& \frac{ K}{(v/D-1 )}
\end{eqnarray}
\begin{figure}[h!]
\begin{center}
{\rotatebox{0}{{\includegraphics[width=0.7\textwidth]{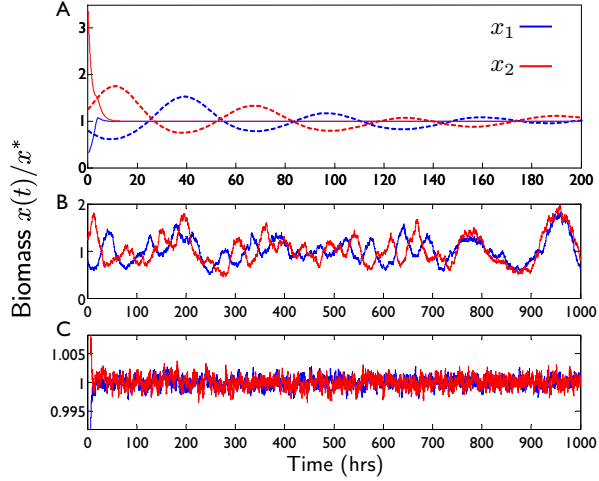}}}}
\caption{Stochastic population oscillations  for the mutualistic cycle model (Fig.~\ref{Fig:cycle}) with $v=1\mathrm{hr}^{-1}$, $D=0.5\mathrm{hr}^{-1}$, $K=1\mu\mathrm{M}$, $\gamma=1.8\times 10^{10}$, $f=0.8$ and volume $V=1$ml. Blue lines correspond to $x_1$ and red to $x_2$. Panel A shows deterministic times series for inflow rate of substrate 1 of $b=0.01\mu$M/hr ($\beta=0.01$ - dashed lines) and of $b=10\mu$M/hr ($\beta=10$ - solid line) calculated by numerical integration of Eqn.\ref{Eq:cycle}. Panels B and C show the equivalent results, but from stochastic simulations, for $b=0.01\mu$M/hr and $b=10\mu$M/hr, respectively. We see that at the lower inflow rate ($b=0.01\mu$M/hr), damped oscillations (A) in the deterministic case become sustained stochastic oscillations in the stochastic model (C). However, for the higher inflow rate ($b=10\mu$M/hr) neither the deterministic (A) nor the stochastic (C) model shows oscillations. These results mirror the observation of stochastically forced oscillations at lower inflow rates in the single population model of Section \ref{sec:stoch}.  Note that the colour codes in B and C are the same as in A.\label{Fig:cyclestraj}}
\end{center}
\end{figure}
Fig.~\ref{Fig:cyclestraj} (panel A) shows deterministic trajectories for the biomass concentrations of microbial populations 1 and 2 predicted by Eqs. (\ref{Eq:cycle}), using the parameter set for {\em{Escherichia coli}} growing on glucose defined in Section \ref{sec:det}, with  $f=0.8$. Results are shown for two different substrate inflow rates. As we observed for the single population in Figure \ref{Fig:DeterministicOsc}, transient oscillations occur for the lower substrate inflow rate  $b=0.01\mu$M/hr; these disappear on increasing the substrate inflow rate to $b=10\mu$M/hr. We hypothesise that the underlying mechanism for these oscillations is the same as that for the single population model: i.e. transient imbalances in the biomass and substrate concentrations which arise because the biomass concentrations change on a comparable or faster timescale than the substrate concentrations.

\begin{figure}[h!]
\begin{center}
{\rotatebox{0}{{\includegraphics[width=0.7\textwidth]{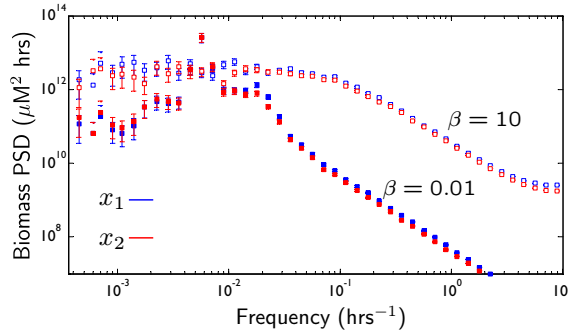}}}}
\caption{Power spectra of the stochastic time series shown in Fig.~\ref{Fig:cyclestraj}C\&D for the mutualistic cycle model. Blue squares correspond to $x_1$ and red to $x_2$; filled symbols correspond to substrate inflow rate $b=0.01\mu$M/hr, while open symbols correspond to $b=10\mu$M/hr. The presence of a peak in the biomass power spectrum for $b=0.01\mu$M/hr  indicates sustained oscillations; no such peak is present for the higher influx rate $b=10\mu$M/hr.
 \label{Fig:MutualPSD}}
\end{center}
\end{figure}

Using the same procedures as in Section \ref{sec:stoch} and \ref{app:kramers-moyal}, we can also carry out stochastic simulations for this model. The resulting trajectories are shown in Fig.~\ref{Fig:cyclestraj}B\&C, for a system volume of $V=1$ml and the same parameters as for the deterministic case. Sustained stochastic oscillations are indeed observed for the substrate inflow rate $b=0.01\mu$M/hr (Panel B) which arise from the transient oscillations in the deterministic model (Fig.~\ref{Fig:cyclestraj}A - dashed lines), while for the higher inflow rate $b=10\mu$M/hr, neither the stochastic (Fig.~\ref{Fig:cyclestraj}C) nor the deterministic system shows oscillations (Fig.~\ref{Fig:cyclestraj}A - solid lines). Fig.~\ref{Fig:MutualPSD} shows biomass concentration power spectra for the stochastic simulations: as expected, for  $b=0.01\mu$M/hr a clear peak is present, indicating sustained stochastic oscillations, while the power spectra show no peak for $b=10\mu$M/hr.

The nutrient-cycling model investigated here represents, of course, only one example of many possible microbial ecosystem topologies. Detailed analysis of the effects of ecosystem topology on the occurrence and character of noise-induced oscillations would be an interesting subject for further work; nevertheless this example suffices to show that the phenomenon is likely to be widespread in microbial ecosystems with small population sizes and unbalanced biomass and substrate removal rates.

\section{Discussion}\label{sec:disc}

In this paper, we have demonstrated that microbial populations can show oscillatory dynamics, under conditions where the biomass is removed from the system faster than the substrate, and the supply rate of substrate is not too high. For large populations these oscillations are transient, and the population eventually reaches a non-oscillatory steady state. For small populations, however, the oscillations can be sustained indefinitely.   The origin of these sustained stochastic oscillations is the  mechanism discovered by McKane and Newman \cite{McKane2005}, in which underlying oscillatory modes of the deterministic system are excited by intrinsic demographic noise. While this mechanism has been shown to produce stochastic oscillations in a number of different systems \cite{McKane2005,Lugo2008,Boland2008,Alonso2007,Ghose2010,Galla2009,Bladon2010,McKane2007,Dauxois2009}, to our knowledge this is the first demonstration of its relevance in a simple model of a single population growing on a substrate. For the single population model, we have used both simulations and analytical arguments to investigate in detail the conditions for the presence of these oscillations and their characteristics. We have also shown using simulations that the same effect can be expected to arise in more complex microbial ecosystems with cross-feeding interactions.

What are the implications of these results for microbial populations in the natural environmnent? Our work suggests that oscillatory dynamics may be widespread in circumstances where microbial populations are small, with high death rates. Small microbial populations are likely to be found in microenvironments such as the interstices between soil grains, on particulate matter in the ocean or, for pathogenic bacteria, the inside of a host cell. In such a closed environment, it is indeed likely that the removal rate of substrate may be low, while the death rate of microbes due to phage predation or attack by host defence mechanisms, may be high. If such oscillations are indeed widespread, this would have important implications for the establishment and maintenance of microbial communities: for example, oscillations would be likely to have a strong effect on the distribution of times to extinction. There may also be an interesting  interplay between microbial ecosystem topology and the characteristics of the stochastic oscillations.

The next stage in this work is clearly to test our predictions experimentally. Traditionally, theories about microbial population dynamics are tested using chemostats, but  in the absence of mortality factors such as phage the conditions required for oscillations are not satisfied in a conventional chemostat, and typical population sizes are anyway almost certainly too large to observe sustained noise-induced oscillations. Microchemostats however, in which small microbial populations are maintained under steady-state conditions in microfluidic devices \cite{Balagadde2005}, may provide conditions under which noise-induced oscillations could be observed. Moreover, microscopic techniques for  observation of the growth of microbes in complex confined geometries \cite{Austin2011} or as biofilm communities \cite{Molin2000},  provide exciting possibilities for testing the likely significance of stochastic oscillations for microbial communities in the natural environment.

\section*{Acknowledgments}
The authors thank Richard Blythe and Mike Cates for valuable discussions and reading of the manuscript, and Billy K. Huang for his contribution to the early stages of this project.  This work was funded by the Leverhulme Trust under grant number F/00158/BX, and by EPSRC under grant number EP/E030173. RJA was funded by the Royal Society of Edinburgh and by a Royal Society University Research Fellowship. AF thanks Prof. Kenneth Murray and the Darwin Trust for fellowship and additional support.

\begin{appendix}

\section{Quality factor of the biomass oscillations}\label{app:qfactor}
The quality factor of a resonant system measures the fidelity of its oscillations; for a mechanical or electrical oscillator, the $Q$ factor can be interpreted as the ratio of the energy stored to energy dissipated over an oscillation cycle \cite{FeynmanLectVol1_Qfactor}. When the $Q$-factor is large an approximate definition is given in terms of its power spectrum $I(\omega)$ as the ratio of the peak oscillation frequency $\Omega_0$ to the range of frequencies that significant oscillations occur $\Delta\omega$: $Q \approx \Omega_0 / \Delta\omega$. A simple damped harmonic oscillator such as a mass on a spring in a viscous fluid, which oscillates at frequency $\omega_0$ with no friction, has the equation of motion:
\begin{equation}\label{Eq:sho}
\frac{\ud^2 x}{\ud t^2} + \frac{\zeta}{m} \frac{\ud x}{\ud t} + \omega_0^2 x = 0
\end{equation}
where $m$ is the mass, $x$ is its position and $\zeta$ is the friction coefficient. For the damped oscillator, the quality factor  $Q$ is proportional to the ratio of energy stored in the spring, to the energy dissipated to frictional loss, per cycle:
\begin{equation}
Q \approx \frac{m \omega_0}{\zeta}
\end{equation}
For the problem under consideration in this paper, our linear approximation of the deterministic dynamics, Eqs (\ref{Eq:jac}) and (\ref{Eq:Jacobian}), allows us to write
\begin{eqnarray}
\frac{\ud \delta x}{\ud t} &\approx& \frac{D\beta}{\gamma}\delta s\\\nonumber
\frac{\ud \delta s}{\ud t} &\approx& -D\gamma \delta x - D (\beta + \chi)\delta s
\end{eqnarray}
which can be combined into a single second-order differential equation for the biomass concentration close to the steady state:
\begin{eqnarray}\label{Eq:biom}
\frac{\ud^2 \delta x}{\ud t^2} + D (\beta + \chi) \frac{\ud \delta x}{\ud t} + D^2 \beta \delta x = 0
\end{eqnarray}
Comparing Eqs (\ref{Eq:sho}) and (\ref{Eq:biom}) we see that our system maps onto the damped simple harmonic oscillator, with $D(\beta + \chi)$ and $D^2 \beta$ playing the roles of $\zeta/m$ and $\omega_0^2$ respectively. We can therefore use this analogy to predict that the quality factor will be given by
\begin{equation}
Q\approx\frac{\sqrt{\beta}}{\beta+\chi}.
\end{equation}

As shown in the main text, the quality factor can be expressed as $Q=\sqrt{\lambda_x/\lambda_s}$, such that the requirement for high $Q$ oscillations is $\lambda_x\gg\lambda_s$. Intuitively, (for small fluctuations around the fixed point) the relaxation rate of the substrate $\lambda_s=\beta+\chi$ is analogous to the rate that a mechanical system dissipates energy $\zeta/m$, and so we can understand the presence or absence of oscillations of biomass and substrate as analogous to the underdamped and overdamped limits of a SHO.

\section{Langevin approximation of the stochastic dynamics}\label{app:kramers-moyal}

In this Appendix, we briefly describe the derivation and implementation of the Langevin approximation of the stochastic model, via a Kramers-Moyal expansion, which we use in our simulations. We follow the procedure described by Van Kampen \cite{vanKampen}. We begin by writing the  Master Equation for the probability $P(n_X, n_S, t)$ of observing the system with $n_X$ microbes and $n_S$ substrate molecules at time $t$. Denoting $(n_X, n_S)$ as the vector $\boldsymbol{n}$, the Master Equation is given by
\begin{equation}\label{Eq:MasterEqn}
\frac{\ud P(\boldsymbol{n},t)}{\ud t}=\sum_{k=1}^4a'_k(\boldsymbol{n}+\boldsymbol{r}_k)P(\boldsymbol{n}+\boldsymbol{r}_k,t)-a'_k(\boldsymbol{n})P(\boldsymbol{n},t)
\end{equation}
In Eq.(\ref{Eq:MasterEqn}), the stoichiometry vector $\boldsymbol{r}_k \equiv (r_{kX},r_{kS})$ denotes the change in $\boldsymbol{n}$ when reaction $k$ fires:  $\boldsymbol{r}_1=(0,1)$, $\boldsymbol{r}_2=(1,-\gamma)$, $\boldsymbol{r}_3=(-1,0)$ and $\boldsymbol{r}_4=(0,-1)$. The propensity function $a'_k$ denotes the probability of occurrence of reaction $k$ per unit time: $a'_1 = b'$, $a'_2 = \mu'(n_S)n_X$, $a'_3 = Dn_X$ and $a'_4 = Rn_S$. The first term in Eq.(\ref{Eq:MasterEqn}) represents the flux of probability from other states  $\boldsymbol{n+r_k}$ into state $\boldsymbol{n}$, while the second term denotes flux of probability due to reactions which move the system out of state  $\boldsymbol{n}$.

We now Taylor expand $a'_k(\boldsymbol{n}+\boldsymbol{r}_k)P(\boldsymbol{n}+\boldsymbol{r}_k,t)$ about $\boldsymbol{n}$, to second order in $\boldsymbol{r}_k$, and substitute the result into Eq.(\ref{Eq:MasterEqn}); this leads to the nonlinear Fokker-Planck equation
\begin{eqnarray}\label{Eq:fpl}
&&\frac{\partial P(\boldsymbol{n},t)}{\partial t} = \sum_{k=1}^4 \left[r_{kB}\frac{\partial q_k}{\partial n_X} +  r_{kS}\frac{\partial q_k}{\partial n_S}\right]\\\nonumber &&+ \frac{1}{2}\sum_{k=1}^4 \left[r_{kX}^2\frac{\partial^2 q_k}{\partial^2 n_X} + 2r_{kX}r_{kS}\frac{\partial^2 q_k}{\partial n_X \partial n_S} + r_{kS}^2\frac{\partial^2 q_k}{\partial^2 n_S}\right]
\end{eqnarray}
where $q_k \equiv a'_k(\boldsymbol{n})P(\boldsymbol{n},t)$. This approximation will be good as long as the number of microbes and substrate molecules is large (see comments in \ref{app:vk}). This in turn can be represented by an equivalent nonlinear  stochastic differential equation (Langevin Equation), for which probability distribution of trajectories generated will follow  Eq.(\ref{Eq:fpl}). This Langevin Equation is given by \cite{vanKampen}:
\begin{equation}\label{Eq:Kramers-Moyal_Langevin}
\frac{\ud\boldsymbol{n}}{\ud t}= \mathsf{A}'(\boldsymbol{n}) +\mathsf{B}'^{1/2}(\boldsymbol{n})\boldsymbol{\xi}'(t)
\end{equation}
where the vector $\mathsf{A}'$ is given by
\begin{equation}\label{Eq:DriftVectorA}
\mathsf{A}' =\sum_{k=1}^4a'_k \boldsymbol{r}_k,
\end{equation}
the matrix $\mathsf{B}'$ is given by
\begin{equation}\label{Eq:VarianceMatrixB}
\mathsf{B}' =\sum_{k=1}^4a'_k \boldsymbol{r}_k\boldsymbol{r}_k^T.
\end{equation}
and $\boldsymbol{\xi}'(t)$ is a vector of Gaussian (white) noise  components, with zero mean and second moment $\langle\boldsymbol{\xi}'\boldsymbol{\xi}'^T\rangle=\mathsf{I}\delta(t-t')$, where $\mathsf{I}$ is the identity matrix. Note that as $\mathsf{B}$ is a symmetric matrix, $\mathsf{V}$ is unitary (i.e. $\mathsf{V}^\dagger=\mathsf{V}^{-1}$), which implies $\mathsf{B}^{1/2}=\mathsf{B}^{1/2\dagger}$. This means multiplying $\mathsf{B}^{1/2}$ by an arbitrary unitary matrix leaves the variance matrix $\mathsf{B}$, and hence, the probability distribution of paths is unchanged, as the Fokker-Planck equation only depends on $\mathsf{B}$. We can transform this Langevin equation in terms of the stochastic dynamics of the concentrations, $x = n_X/V$ and $s = n_S/V$, by dividing through by the volume of the system $V$. The result is that the propensities $a_k=a'_k/V$, so that $a_1=b$, $a_2 = \mu(s)x$, $a_3 = Dx$ and $a_4 = Rs$, leading to the Langevin equation in the main text (Eqn.\ref{Eq:Kramers-Moyal_Langevin_conc}), with $\mathsf{A}$ and $\mathsf{B}$ given by Eqns  \ref{Eq:AVector_conc} and \ref{Eq:Bmatrix_conc} and where the noise term $\xi=\xi'/\sqrt{V}$.

In our simulations, we compute at each timestep the propensities $a_k$ for each reaction $k$, and hence obtain the vector  $\mathsf{A}$ and matrix $\mathsf{B}$. We then compute the square root of $\mathsf{B}$ numerically, by calculating the eigenvalue decomposition of $\mathsf{B}=\mathsf{VDV}^\dagger$, to give $\mathsf{B}^{1/2}=\mathsf{VD}^{1/2}\mathsf{V}^{\dagger}$, where $\mathsf{D}$ is the diagonal matrix of eigenvalues and $\mathsf{V}$ is a matrix whose columns are the eigenvectors of $\mathsf{B}$ and $\dagger$ is the transpose and conjugate operation (Hermitian conjugate).  The components of the vector $\boldsymbol{\xi}$ are obtained using a random number generator. This allows us to update the microbe and substrate populations using an Euler integration scheme with timestep of 0.05hrs.

\section{Analytical calculation of the power spectrum}\label{app:vk}
To obtain analytical expressions for the power spectrum of biomass and substrate concentration fluctuations, we need to find a way of expressing dynamics of the fluctuations around the fixed point. To do this we apply van Kampen's large system expansion method \cite{vanKampen} to the Master Equation (\ref{Eq:MasterEqn}).  This method relies on the ansatz that  the solution of  Eq.(\ref{Eq:MasterEqn}) will be of the form
\begin{equation}\label{Eq:van_KampenAnsatz}
\boldsymbol{n(t)}=V\boldsymbol{\phi}_0(t)+\sqrt{V}\boldsymbol{\psi}(t)
\end{equation}
The vector function $\boldsymbol{\phi}_0(t)$ is the solution of the deterministic equations (\ref{Eq:ndot}) and (\ref{Eq:sdot}) or equivalently
\begin{equation}\label{Eq:MacroscopicPart}
\ud \boldsymbol{\phi}_0/\ud t=\mathsf{A},
\end{equation}
while the vector function $\boldsymbol{\psi}(t)$ represents the fluctuations around the deterministic solution, caused by noise, which we expect to be of size $\sqrt{V}$. The Van Kampen method then formally expands the Master eqn in orders of $V^{-1/2}$ about the deterministic solution to give a linear noise Fokker Planck or Langevin equation. We can do this by starting with the Langevin equation from the Kramers-Moyal expansion (Eqn.\ref{Eq:Kramers-Moyal_Langevin}) and substituting in Eqn.\ref{Eq:van_KampenAnsatz} on both sides of the equation to give:
\begin{equation}\label{Eq:Kramers-Moyal_Langevin_VK}
\begin{split}
V\frac{\ud \boldsymbol{\phi}_0}{\ud t}+\sqrt{V}\frac{\ud\boldsymbol{\psi}}{\ud t}= \mathsf{A}'(V\boldsymbol{\phi}_0&+\sqrt{V}\boldsymbol{\psi})\\ &+\mathsf{B}'^{1/2}(V\boldsymbol{\phi}_0+\sqrt{V}\boldsymbol{\psi})\boldsymbol{\xi}'(t)
\end{split}
\end{equation}
Each of the $\mathsf{A}'$ and $\mathsf{B}'$ are composed of terms $a'_k\boldsymbol{r}_k$, a Taylor expansion of which gives
\begin{equation}\label{Eq:TaylorExpansionPropensity}
\begin{split}
a'_k(V\boldsymbol{\phi}_0+\sqrt{V}\boldsymbol{\psi})&\approx a'_k(V\boldsymbol{\phi}_0)\boldsymbol{r}_k+ \sqrt{V}\boldsymbol{r}_k(\boldsymbol{\nabla}a'_k\cdot\boldsymbol{\psi})\\
&=a'_k(V\boldsymbol{\phi}_0)\boldsymbol{r}_k+ \sqrt{V}\mathsf{J}_k\boldsymbol{\psi}
\end{split}
\end{equation}
where the contribution to the Jacobian of each reaction is given by $\mathsf{J}_k=\boldsymbol{r}_k(\boldsymbol{\nabla} a'_k)^T$; these are related to the Jacobian by $\mathsf{J}=\sum_k\mathsf{J}_k$. Using Eqn.\ref{Eq:TaylorExpansionPropensity} in Eqn.\ref{Eq:Kramers-Moyal_Langevin_VK} and keeping only terms of order $V$ and $\sqrt{V}$ and assuming that $V$ is large so we only retain terms linear in the noise (known as the Linear Noise Approximation), we have
\begin{equation}\label{Eq:Kramers-Moyal_Langevin_VK}
\begin{split}
V\frac{\ud \boldsymbol{\phi}_0}{\ud t}+\sqrt{V}\frac{\ud\boldsymbol{\psi}}{\ud t}= V\mathsf{A}(\boldsymbol{\phi}_0)&+\sqrt{V}\mathsf{J}\boldsymbol{\psi}\\ &+\sqrt{V}\mathsf{B}^{1/2}(\boldsymbol{\phi}_0)\boldsymbol{\xi}'(t),
\end{split}
\end{equation}
where we have used the fact that $a'_k(V\boldsymbol{\phi}_0)=Va_k(\boldsymbol{\phi}_0)$. Further, using Eqn.\ref{Eq:MacroscopicPart} means we can cancel the terms of order $V$ that relate to the deterministic or `macroscopic' evolution of the system to leave a Langevin Equation for the stochastic part:
\begin{equation}
\frac{\ud\boldsymbol{\psi}}{\ud t}=\mathsf{J}\boldsymbol{\psi}+\mathsf{B}^{1/2}(\boldsymbol{\phi}_0)\xi'(t).
\end{equation}
However, we are interested in the fluctuations in concentration about the steady state or fixed point, so we can form the variable $\boldsymbol{z}=\boldsymbol{\psi}/\sqrt{V}=(\delta x,\delta s)^T$ and consider the solution for $\boldsymbol{\phi}_0(t\rightarrow\infty)=\boldsymbol{\phi}^*$ to give an effective linear-noise Langevin equation for the fluctuations around the steady state:

\begin{equation}\label{Eq:Langevin}
\frac{\ud\boldsymbol{z}}{\ud t}= \mathsf{J}^*\boldsymbol{z}+\mathsf{B}^{*1/2}\boldsymbol{\xi}(t)
\end{equation}

\noindent where $\mathsf{J}^*$ is the Jacobian matrix of the dynamics around the fixed point, given in Eqn.\ref{Eq:Jacobian} and the variance matrix $\mathsf{B}^*=\mathsf{B}(\boldsymbol{\phi}_0^*)$ of the effective diffusion process is Eqn.\ref{Eq:VarianceMatrixB} evaluated at the fixed point:
\begin{equation}
\begin{split}
\mathsf{B}^* & =\sum_{k=1}^3\lambda_k \boldsymbol{r}_k\boldsymbol{r}_k^T\big|_{\boldsymbol{x}=\boldsymbol{x}^*}\\
&=(b-Rs^*)\left(
             \begin{array}{cc}
               2/\gamma & -1 \\
               -1  & \frac{b+Rs^*}{b-Rs^*}+\gamma \\
             \end{array}
           \right)
\end{split}
\end{equation}
The Langevin noise term is Gaussian with zero mean and moment $\langle\boldsymbol{\xi}\boldsymbol{\xi}^T\rangle=\mathsf{I}\delta(t-t')/V$. The van Kampen and Kramers-Moyal approaches produce linear and non-linear Fokker-Plank or stochastic differential equations. However, as we have shown they can be obtained from each other by a change of variable and are equivalent to within fluctuations of order $\sqrt{V}$, which in any case is the level of approximation of each approach.

The power spectrum of fluctuations around the steady state solution can be calculated from the Fourier Transform (FT) of Eqn \ref{Eq:Langevin} as

\begin{equation}
\langle \boldsymbol{Z}(\omega)\boldsymbol{Z}(\omega)^\dagger\rangle = \frac{1}{V}\mathsf{K}(\omega)\mathsf{B}^*\mathsf{K}^\dagger(\omega)\\
\end{equation}

\noindent where $\boldsymbol{Z}(\omega)=\mathrm{FT}\{\boldsymbol{z}(t)\}$ and $\mathsf{K}(\omega)=(i\omega\mathsf{I}-\mathsf{J})^{-1}$. In particular, if we assume that $\gamma\gg1$ then we find the power spectrum of biomass and substrate fluctuations are as given by Eqns. \ref{Eq:nPSD} \& \ref{Eq:sPSD} in the main text.

\end{appendix}

\bibliography{KhatriBibliography}
\bibliographystyle{model1a-num-names}

\end{document}